\documentclass[doublecol]{epl2} 
\usepackage[]{amsfonts}
\usepackage[]{amssymb}
\usepackage[]{bm}
\usepackage[]{mathrsfs}

        \title{Self-organizing dynamical networks able to learn autonomously}

        \author{Pablo Kaluza}

        \institute{Interdisciplinary Institute of Basic Sciences, National Scientific and Technical Research Council (CONICET) \& Faculty of Exact and Natural Sciences, National University of Cuyo. Padre Contreras 1300, 5500 Mendoza, Argentina.}
        
        \pacs{89.75.Hc}{Networks and genealogical trees}
        \pacs{05.65.+b}{Self-organized systems}
        \pacs{05.45.-a}{Nonlinear dynamics and chaos} 
 	
        \abstract{We present a model for the time evolution of network architectures based on dynamical systems. We show that the evolution of the existence of a connection in a network can be described as a stochastic non-markovian telegraphic signal (NMTS). Such signal is formulated in two ways: as an algorithm and as the result of a system of differential equations. The autonomous learning conjecture [Phys. Rev. E \textbf{90},030901(R) (2014)] is implemented in the proposed dynamics. As a result, we construct self-organizing dynamical systems (networks) able to modify their structures in order to learn prescribed target functionalities. This theory is applied to two systems: the flow processing networks with time-programmed responses, and a system of first-order chemical reactions. In both cases, we show examples of the evolution and a statistical analysis of the obtained functional networks with respect to the model parameters.}

\begin{document}

    \maketitle

    \section{Introduction}
	
	Dynamical systems that model complex networks are characterized by dynamics that can operate on fixed network architectures \cite{Vespignani} or on dynamical networks \cite{Strogatz,Dorogovtsev1,Gross} where the evolution of the network architectures is part of the dynamical system. The second approach can be seen as the general case and it is one of the main research topics in the field of complex networks \cite{Dorogovtsev2,Barabasi_networks}. The close relationship between the dynamics on the network and the network architecture indicates that these two entities cannot be studied separately \cite{nishikawa}.
	
	Our goal is to construct dynamical systems that evolve on networks that also feature dynamics. These systems must be functional, i. e., they must have operational functionalities that can be quantitatively measured. In addition, we require them to learn how to set their parameters autonomously in order to reach the prescribed functionality. Particularly in this work, the parameters that must be learned concern the existence of connections between nodes.
	
	The problem of constructing dynamical networks with prescribed functionalities has been developed extensively by applying a supervised learning formalism \cite{machine_learning}. Genetic networks with adaptive responses have been constructed by a genetic algorithm \cite{Inoue-Kaneko} and by annealing optimization \cite{kaluza-inoue}. Functional and robust flow processing networks have been established \cite{Kaluza_PRE_2007,Kaluza_EPL_2007,Kaluza_CHAOS_2008,kaluza_kori_mikhalov_2008,Kaluza_EPJB_2012,Kaluza_EPJB_2017_coe,Hutt}. Genetic networks with prescribed oscillatory properties have been also considered  \cite{Yasuaki1,Yasuaki2}. Finally, we can mention systems of phase oscillators that are constructed to perform required levels of synchronization \cite{Yanagita1,Yanagita2,Zoran}.

	Control theory has also been employed to obtain target dynamics of complex systems \cite{barabasi_control_2016}. Much work has been done in delayed feedback schemes of control \cite{Scholl}. An important study in this field was conducted by Pyragas \cite{pyragas}. In this case, a dynamical system is controlled by a feedback loop that helps stabilizing unstable orbits. Note that this feedback does not change the parameters of the system, but adds a new control term to the dynamics.
	
	Our approach is different from previous ones in that we expect the dynamical system to incorporate the learning capacity in their own dynamics. Thus, this system can learn autonomously, in a self-organized way, to reach a desired functionality. The main goal of this scheme is for the dynamical system to learn without the intervention of an external tutor. Already we have shown in previous works that this concept can be applied to different dynamical systems such as phase oscillators \cite{kaluza_autonomous}, feed-forward neural networks \cite{kaluza_autonomous_discrete}, and for gradient dynamics and control of oscillation death in Kuramoto-Tsuzuki systems of oscillators \cite{kaluza_autonomous_perdido}.
	
	In this work, we design link dynamics based on our previous conjecture of autonomous learning. As a result, the designed dynamics operate analogously to a telegraphic random signal.  The transitions between the existence and absence of a link is controlled by the previous state and a delayed state of the system (memory).	In addition to presenting the algorithmic scheme of this kind of telegraphic signal, we show how to generate it from basic quantities of the system. Finally, we apply these concepts to two kind of networks of different nature: flow-processing networks with discrete time dynamics, and chemical systems of first-order reactions with continuous time dynamics. For these two examples, we present statistical results of autonomous learning performance as well.
	
	This work is organized as following. In the next section, we present the main concepts of autonomous learning, the definition of the link dynamics as a telegraphic variable, and how to generate this random variable by hybrid dynamics. In the second part of this manuscript, we show two examples of networks. Finally, we present the discussion and the conclusion in the last section.

	\section{Model}
	
	We consider a dynamical system $G$ constituted by $N$ nodes which are connected according to an adjacency matrix $\mathbb{A}$. If there exists a link from node $i$ to node $j$, $A_{ji}=1$; otherwise $A_{ji}=0$. Each node $i$ is characterized by a dynamical variable $x_i$. The dynamics of the system $G$ are expressed as
	
	\begin{equation}
            \boldsymbol{\dot{x}} = \boldsymbol{f}(\boldsymbol{x}, \mathbb{A}).
            \label{equ_dynamics}
	\end{equation}

	The task performed by the dynamical system $G$ is defined by a functional $\boldsymbol{F}(\boldsymbol{x})$. The error $\epsilon(G)$ of this system with respect to a target task $\boldsymbol{R}$ is 
	
	\begin{equation}
            \epsilon(G) = || \boldsymbol{F}(\boldsymbol{x}) - \boldsymbol{R}||.
            \label{equ_error}
	\end{equation}
	
	\noindent
	A functional system must be able to find an adjacency matrix $\mathbb{A}_R$ that minimizes the error $\epsilon(G)$. We propose to extend the system (\ref{equ_dynamics}) to a new one where the links $A_{ij}$ have their dynamics based on our autonomous learning conjecture. 
	
        In essence, the problem we present here is not different from the one that we have considered in our previous work about autonomous learning \cite{kaluza_autonomous}. In effect, the links of the adjacency matrix play the role of the parameters $\boldsymbol{w}$ in our previous formulation. However, a fundamental difference arises here. The parameters $\boldsymbol{w}$ are continuous variables with continuous time dynamics. Now, the links are discrete variables with continuous or discrete time dynamics. This mayor difference does not allow the direct application of the previous method of autonomous learning, so a new one is developed here.

        \subsection{Link dynamics as a non-markovian telegraphic signal (NMTS)}
        
        The dynamics of a link in a network can be seen as a sequence of ones and zeros that represent the existence of that connection between two nodes. This discrete sequence of states can be continuous or discrete on time. It results natural to model these dynamics with a random variable similar to a telegraphic signal \cite{telegraphi_signal_properties}. Thus, the main work is to relate the transitions between the two possible states with the current state and performance of the system, and, these values to a past instant of it. 
    
        We consider that during the evolution of the dynamical system (\ref{equ_dynamics}), the link states are updated synchronously at fixed time intervals $\tau$. Thus, the link states present a discrete time evolution. We call these iterations \textit{epochs} of their dynamics. If the dynamical system (\ref{equ_dynamics}) has a unity time scale, we require that $\tau \gg 1$ to ensure that the system can define the error function properly (\ref{equ_error}). 
    
        Consider a system $G(t)$ with an adjacency matrix $\mathbb{A}(t)$ and an error $\epsilon(t)$. 
        In a past instant at time $t-\Delta$, the system $G(t-\Delta)$ has an adjacency matrix $\mathbb{A}(t-\Delta)$ and an error $\epsilon(t-\Delta)$. 
        Here $\Delta$ plays the role of a delay with an integer number $\eta$ of $\tau$ intervals, i. e., $\Delta = \eta \tau$.
        We compute the difference for each possible link  $\delta A_{ij} = A_{ij}(t) - A_{ij}(t-\Delta)$, and for the errors of the systems $\delta \epsilon = \epsilon(t)- \epsilon(t-\Delta)$.  Thus, if $\delta A_{ij} = 1$ a new links has been created, whereas if $\delta A_{ij} = -1$, the link has been removed from the network. In the case of $\delta A_{ij}=0$, there is no change in this link configuration. The difference $\delta \epsilon < 0$ indicates that the system  improves its performance, while $\delta \epsilon > 0$ signals its degradation with respect to the past configuration. Similarly, if $\delta \epsilon = 0$ there is not a change in the system performance. 
        
		\begin{table}[!ht]
	\begin{center}
	\begin{tabular}{ | c | c | c  | c | c |}
			\hline
			case & $\delta A_{ij}$ & $\delta \epsilon$ & $\delta A_{ij} \delta \epsilon$ & $A_{ij}(t')$ \\ 
			\hline
			1 & 1 & + & + &  0\\ 
			\hline
			2 & 1 & - & - &  $A_{ij}(t)$\\ 
			\hline
			3 & -1 & + & - &  1\\ 
			\hline
			4 & -1 & - & + &  $A_{ij}(t)$\\ 
			\hline
			5.a  & $\pm1$  & $0$  & 0 & transition rate $q = S\epsilon$ \\
			5.b  & $0$  & $\pm$  &   & $1$ if $A_{ij}(t)=0$\\
                        5.c  &  $0$ & $0$  &   & $0$ if $A_{ij}(t)=1$\\
			\hline 
		\end{tabular}
		\caption{Algorithmic description of the stochastic non-markovian telegraphic signal (NMTS). It shows all possible cases of $\delta A_{ij}$ and $\delta \epsilon$.}
		\label{table_options}
		\end{center}
		\end{table}
	
	By taking these variations into account, we can use the autonomous learning conjecture to propose the needed changes in the network for reducing the error at the next epoch $t' = t + \tau$. If a link has been added ($\delta A_{ij}=1$) and the error increases ($\delta \epsilon>0$), we must revert the addition of this link making $A_{ij}(t') =0$. This situation corresponds to the first case in table \ref{table_options}. The second case of the table indicates that a link has been added, and the error decreases. Thus, we keep the link addition by making $A_{ij}(t') = A_{ij}(t)$. Cases three and four are analogous to the first two but when a link has been removed.
	
	The fifth case in table \ref{table_options} presents the particular situation when $\delta A_{ij} \delta \epsilon = 0$. This instance can occur when two different configurations have the same error (5.a), or the same existence value for the link $A_{ij}(t) = A_{ij}(t-\Delta)$ (5.b), or both (5.c). Thus, we use this last case to introduce a kind of mutation. We say that we can change the configuration of the system with a probability $q$. That is, if $A_{ij}(t) = 0$, then $A_{ij}(t') = 1$, and if $A_{ij}(t) = 1$, then $A_{ij}(t') = 0$. Note that this probability of mutation must decrease when the system is reaching the target function. We define $q = S\epsilon$, where $S$ is the noise intensity parameter. This way, the changes become smaller when the error decreases. Note that this scheme tries to mimic an annealing-like algorithm \cite{annealing}.
	
	As a result of our choices, the dynamics of a link can be written as a non-markovian telegraphic signal (NMTS) $L$ in which the transitions between states are functions of the current and past configuration (memory) of the system and their performance:
	
        \begin{equation}
		A_{ij}(l+1) = L \big(A_{ij}(l),A_{ij}(l - \eta),\epsilon(l), \epsilon(l - \eta) \big).
		\label{link_evo_continua}
	\end{equation}

	\noindent
	This expression is formulated in epochs. Our conjecture is that the new system consisting of eqs. (\ref{equ_dynamics}) and (\ref{link_evo_continua}) evolves decreasing the error (\ref{equ_error}).

        \subsection{Generation of the non-markovian telegraphic signal}
        
        In the previous subsection we showed that a kind of telegraphic signal is able to produce dynamics for the connectivity of a network. However, the previous formulation is far from our goal of autonomous learning. In effect, such formulation is given in an algorithmic way, i. e., we need a tutor that applies an algorithm. We propose now to show that it is possible to extend the original dynamical system (\ref{equ_dynamics}) defining new dynamical variables that incorporate the autonomous learning scheme.
        
        We can apply a version of the autonomous learning for each possible link in terms of epoch as follows
    
        \begin{eqnarray}
		\mathscr{A}_{ij}(l+1) &=&  A_{ij}(l) \nonumber \\
                                &-& K \Big( A_{ij}(l) - A_{ij}(l-\eta) \Big) \Big( \epsilon(l) - \epsilon(l-\eta) \Big) \nonumber \\
                                &+& f(S,\epsilon) \xi_{ij}(l)
		\label{equ_autonomo_link1}
	\end{eqnarray}
	
	and
	
	\begin{equation}
		A_{ij}(l+1) = \Theta \Big(\mathscr{A}_{ij}(l+1) - h \Big).
		\label{equ_autonomo_link2}
	\end{equation}

	\noindent
	Here, $\Theta(x)$ is the Heaviside step function, with $\Theta(x)=0$ if $x \le 0$ and $\Theta(x)=1$ if $x > 0$. The parameter $h$ is a threshold that defines whether a link exists. The stochastic variable $\xi_{ij}(t)$ is white noise, with $\langle \xi_{ij}(t) \rangle = 0$ and $\langle \xi_{ij}(t) \xi_{kl}(t') \rangle = 2 \delta_{ik} \delta_{jl} \delta_{tt'}$. The function $f(S, \epsilon)$ controls noise intensity, with $S$ acting as the intensity parameter. The parameter $K$ indicates the importance of the drift term of the dynamics.
	
	Equation (\ref{equ_autonomo_link1}) is a kind of discrete version of the one that we have previously introduced for continuous parameters \cite{kaluza_autonomous}. However, it presents an important new characteristic. The right side of this equation uses the link states $A_{ij}(t)$ and $A_{ij}(t-\Delta)$ as discrete variables with possible values zero and one. The left side of the equation, however, introduces a new variable $\mathscr{A}_{ij}$ that has a real domain. Formally, eq. (\ref{equ_autonomo_link1}) and eq. (\ref{equ_autonomo_link2}) define an hybrid dynamical systems where the dynamics feature jumps for certain conditions \cite{SD_hibridos}.
	
	Equation (\ref{equ_autonomo_link2}) transforms the continuous variable $\mathscr{A}_{ij}(t)$ back into a discrete one. Note that the variable $\mathscr{A}_{ij}$ is introduced here only to help formulating the system. However, it is not a hidden weight for the interaction between nodes  $i$ and $j$. The parameter $S$ plays the same role as in table \ref{table_options}.
	
	In general, we take $K \gg 1$. The effect of this choice is simple to understand. The product $\delta A_{ij} \delta \epsilon$ operates in the same way as in table \ref{table_options}. As a result, in the next epoch we add that correction to the previous instant $A_{ij}(l)$. Since $K$ is large, the correction is strong and the variable $\mathscr{A}_{ij}(t)$ takes very large positive or negative values far from the threshold $h$. Then, eq. (\ref{equ_autonomo_link2}) generates a discrete value based on these corrections. In these cases, for realistic values of the noise intensity $f(S,\epsilon)$, the stochastic dynamics plays no role on these corrections.
	
	The situation is different when $\delta A_{ij} \delta \epsilon = 0$. Now, the stochastic dynamics drive the system. It is clear that the noise adds a random value to the previous state $A_{ij}(l)$. The probability that this result crosses the threshold $h$ depends on the noise intensity $f(S, \epsilon)$. Then, by choosing this function properly, we can produce the same effect of random mutation as in the algorithmic version (NMTS).
	
	In conclusion, our subsystem of eqs. (\ref{equ_autonomo_link1}) and (\ref{equ_autonomo_link2}) produces the same results as the NMTS defined in table \ref{table_options}. However, it is important to understand the significant differences between the use of a an algorithm to create the stochastic signal and the autonomous way for its generation by a map.

	\subsection{The noise intensity function $f(S, \epsilon)$}
	
	In the previous subsection we have discussed how the transitions and mutations in our scheme works with eqs. (\ref{equ_autonomo_link1}) and (\ref{equ_autonomo_link2}). We observe that these transitions have a strong dependency on the noise intensity function $f(S, \epsilon)$. It is clear that the probability $q(S, \epsilon)$ to cross the threshold $h$ is given by
	
	\begin{equation}
            q(S,\epsilon) = \frac{1}{2}\Bigg[1- erf \Bigg( \frac{h}{\sqrt{2} f(S, \epsilon)} \Bigg) \Bigg].
            \label{equ_p_de_S_e}
	\end{equation}
    
        Our particular choice is to make $q(S, \epsilon) = S\epsilon$ in order to reproduce the algorithm of table \ref{table_options}. It is straightforward to show that the function $f(S, \epsilon)$ must be taken as
	
	\begin{equation}
            f(S,\epsilon) = \frac{h}{\sqrt{2} erf^{-1}(1-2 S \epsilon)}. 
	\end{equation}
	
	\noindent
	Here, $erf^{-1}(x)$ is the inverse of the error function $erf(x)$. As a result, the probability of transitions per epoch is the same as we defined in our algorithmic scheme with the NMTS. In this work we take always $h = 0.5$. Note that $f(S, \epsilon) \to 0$ when $\epsilon \to 0$.

	\section{Numerical results} 
	
	We present two systems where we can apply our new formulation. The first one is a model of flow processing network, and the second one, is a network of first order chemical reactions.
	
	\subsection{Flow processing networks}
	
	The first example we consider is the flow processing network model with time-programmed responses. This has been studied by the author in prior works \cite{Kaluza_EPJB_2012} and \cite{Kaluza_EPJB_2017_coe}. In those articles, we have used an annealing-like algorithm to perform supervised learning in order to obtain functional networks. Definitions and properties of this model can be found from the previous references.
	
	\begin{figure}[!ht] 
		\begin{center}
			\includegraphics[width=0.9\columnwidth,clip]{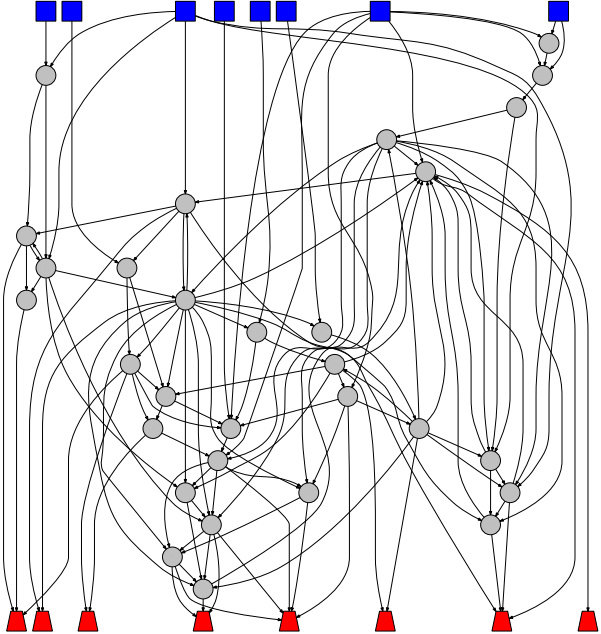} 
			\caption{Final network at the end of the evolution shown in fig. \ref{fig_evolution}. The network has $N=46$ elements divided into $N_{in}=8$ input nodes (blue squares), $M=30$ middle nodes (gray circles) and $N_{out}=8$ output nodes (red trapeziums).}
			\label{fig_net} 
		\end{center}
	\end{figure}
	
	A network of this model presents a layered structure with $N$ nodes divided into $N_{in}$ input nodes, $M$ middle nodes and $N_{out}$ output nodes. Input nodes receive fluxes that are redistributed by the middle nodes. The output nodes obtain the final fluxes as responses. Note that a functional network must be able to produce a desired set of flux levels in the output nodes (target response).  Figure \ref{fig_net} presents an example of these networks. The difference between the actual output pattern and the target one is called the flow error $\epsilon$ of a network. These networks are characterized by their discrete-time dynamics; thus, the error of a network with respect to a target function is not defined for all time, and we need to wait for some iterations until the network can generate its response. As a result, we consider that the link dynamics operate after the network can produce its response.
	
	\begin{figure}[!ht] 
		\begin{center}
			\includegraphics[width=1.0\columnwidth,clip]{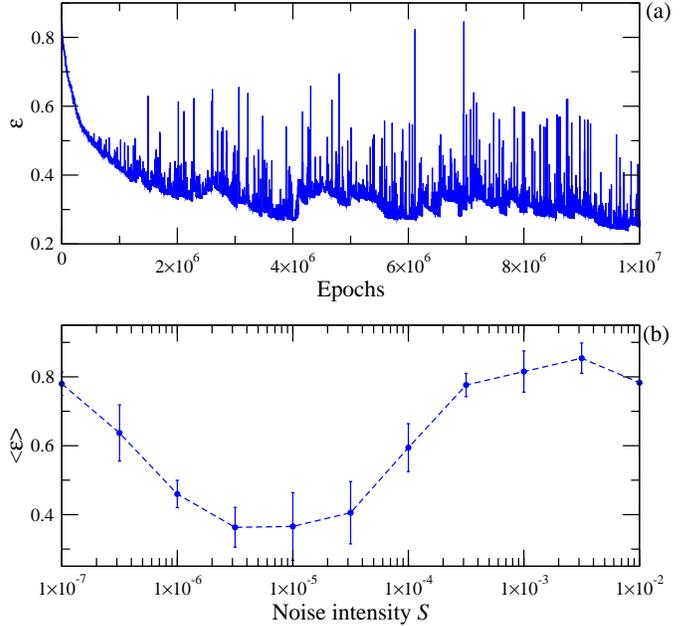} 
			\caption{(a) Flow error $\epsilon$ as a function of the number of epochs for one network evolution with $N=46$ nodes. (b) Mean flow error as a function of the noise intensity $S$. Each point corresponds to the average value over $100$ realizations. Error bars indicated the dispersion of the distributions of the realizations for each ensemble.}
			\label{fig_evolution} 
		\end{center}
	\end{figure}
	
	We present a typical evolution pattern for one of these networks in fig. \ref{fig_evolution}a from the link dynamics given by eqs. (\ref{equ_autonomo_link1}) and (\ref{equ_autonomo_link2}). As the first example, we consider network with $N=46$ elements divided into $N_{in}=8$ input nodes, $M=30$ middle nodes and $N_{out}=8$ output nodes, with noise intensity $S = 1 \times 10^{-4}$ and $K=10^4$. The initial conditions for the evolution take a random structure with connectivity $p=0.2$. This network is also taken as memory for the first epoch delay ($\eta = 1$). As a result, the initial errors for the actual network and its memory are the same.
	
	We find that the flow error $\epsilon$ decreases with the number of epochs, and the error value becomes relatively small at the end of the evolution. Note that networks with this size, constructed in \cite{Kaluza_EPJB_2012} by using an annealing algorithm, could reach a mean error $\langle \epsilon \rangle \approx 0.25$ at best. Thus, our new autonomous method can reach similar performance for the chosen number of epochs\footnote{In the present work we change the normalization of the error definition used in \cite{Kaluza_EPJB_2012}. Actual error $\epsilon$ must be divided by $2N_{in}T = 2\times 8 \times 10 = 160$ to get consistent values of error with the ones shown in \cite{Kaluza_EPJB_2012}.}. Figure \ref{fig_net} presents the final network from the evolution shown in fig. \ref{fig_evolution}a. We can observe during the evolution that at times, large values of flow error appear, but normally they are corrected in the following epochs. This is because the network performs each time better and a random mutation has in general a negative effect on the network functionality (larger $\epsilon$). 
	
	In the second part of the first example, we study the performance of the evolutions with respect to the noise intensity $S$. We consider several ensembles of $100$ networks that evolve with the same value of $S$. All simulations have $10^7$ epochs. Each evolution in an ensemble reproduces a random target pattern; therefore, the outcome tends to an average over all possible functionalities. Figure \ref{fig_evolution}b presents the mean flow errors $\langle \epsilon \rangle$ and their dispersions for several ensembles as a function of the noise intensity parameter $S$. We observe that this curve has a minimum for $S=3.12 \times 10^{-4}$. For larger and smaller values of $S$, the mean flow error increases. The maximum and minimum values in these simulations are compatible with those found in our previous works \cite{Kaluza_EPJB_2012,Kaluza_EPJB_2017_coe}. 
	
	Finally, it is interesting to make a comparison between the results of this work and those obtained from an annealing algorithm in \cite{Kaluza_EPJB_2012} and \cite{Kaluza_EPJB_2017_coe}. Although the networks constructed by these two different methods are at the end quite similar, the evolutions with our new dynamics are slower in terms of epochs. We need nearly one order of magnitude more epochs to find networks with similar performance (flow error). There are two main reasons for this difference. The first is that the new dynamics always accept non-functional mutations that are removed in the next epochs. As a result, twice as many iterations are required to deal with this effect. The second reason is that there is one link mutation per epoch in the annealing algorithm, whereas all links can change simultaneously in the new autonomous dynamics.

	\subsection{First order chemical reaction networks}
	
\begin{figure}[!ht] 
		\begin{center}
			\includegraphics[width=1.0\columnwidth,clip]{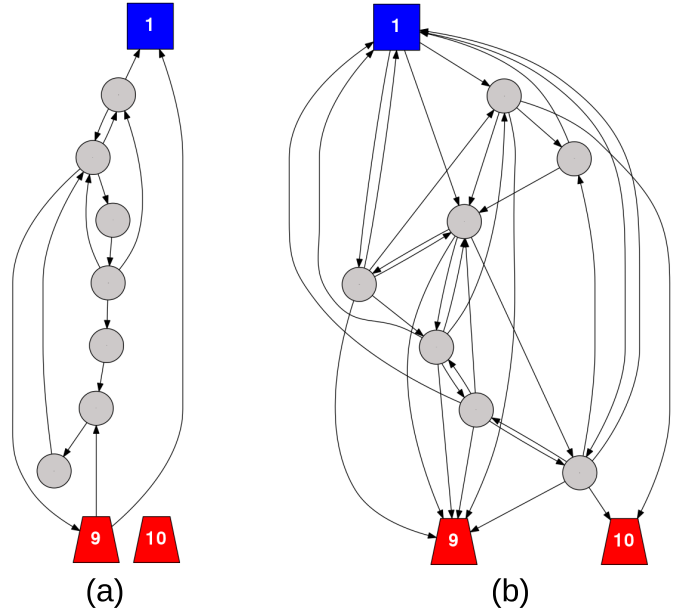} 
			\caption{Chemical networks. Initial network (a) and final network (b) of the evolution shown in fig. \ref{fig_evo_quimico}. Networks have $N=10$ nodes with one input node (blue square) and two output nodes (red trapeziums).}
			\label{fig_nets} 
		\end{center}
	\end{figure}

	As a second example, we develop networks of first order chemical reactions. They are characterized by linear dynamics for the concentrations of the compounds involved \cite{chemical_reactions}. A system of these chemical reactions is described by an adjacency matrix $\mathbb{A}$ whose elements $A_{ij}=1$ if compound $j$ produces compound $i$, and $A_{ij}=0$ otherwise. Self-connections are not allowed. Figure \ref{fig_nets} shows networks with this architecture. The dynamics of the compound concentration $x_i$ of a system with $N$ nodes is governed by following system of differential equations
	
	\begin{equation}
            \dot{x}_i = \sum_{k=1}^N A_{ik}x_k  - \sum_{k=1}^N A_{ki} x_i- \gamma x_i + I^{ext}_i.
            \label{net_quimica_dinamica}
	\end{equation}
	
	\noindent
	In this expression, $\gamma$ is the rate of degradation of the chemical species and $I_{i}^{ext}$ is an external input flow applied to the node $i$. Note that all the kinetics rate constants between compounds are equal to the unity in this model.
	
	These dynamics imply that the total instantaneous concentration evolves to a constant value $C = \sum_{k=1}^N I_k^{ext}/\gamma = \sum_{k=1}^N x_k(t)$ when ($t \to \infty$). The network dynamics settle in a stable fixed point in the space of concentrations. Thus, the concentration of each compound in the stable fixed point depends on the network architecture $\mathbb{A}$ of the reactions.

	The purpose of a network in this model is to establish certain values of concentrations $\hat{x}_i$ for the last $N_{out}$ nodes of the network. We consider that only an external input flow is applied to the first node of the network, i. e,  $I^{ext}_1 \ne 0$. Thus, the reactions propagate from this input node according to the network structure $\mathbb{A}$.

	We define the error $\epsilon(t)$ of the system as the sum of the absolute differences between the current and target values of concentrations:     
	
	\begin{equation}
            \epsilon(t) = \frac{1}{2C} \sum_{i=N-N_{out}+1}^{N} |x_i(t) - \hat{x}_i|.
            \label{equ_error_reactions}
	\end{equation}
	
	\noindent
	We can summarize the problem now: given a set of target concentrations $\{\hat{x}_i\}$, the network architecture must evolve by eqs. (\ref{equ_autonomo_link1}) and (\ref{equ_autonomo_link2}) minimizing the error (\ref{equ_error_reactions}), whereas the concentrations evolve according with (\ref{net_quimica_dinamica}).

	The first example presented for this system is the evolution of a network of $N=10$ elements and with $N_{out}=2$ nodes. The target values required for the output nodes are $\hat{x}_{9} = 6.5$ and $\hat{x}_{10}=2.5$, while for the external input flow, $I_1^{ext}=1$. The degradation rate is $\gamma= 0.1$, hence, the total concentration is $C=10$. Note that we require only that the $90\%$ of the total concentration $C$ arrives to the output nodes to avoid unattainable target concentrations given the rate constants of the reactions.  The other parameters of the system are $S=0.005$ and $\eta =1$. Since the characteristic time of this system is given by $1/\gamma = 10$, we take $\tau = 20$ ($\Delta = 20$). Thus, we ensure that the system has enough time to define properly the error $\epsilon$ between epochs. Numerical integration is done with a second-order Runge-Kutta algorithm with $dt=0.01$. The initial random network has a probability of connection of $p=0.2$. Initial concentrations are $x_i = 0$. The delayed system (memory) is initialized with the same network structure and error of such initial network.

	\begin{figure}[!ht] 
		\begin{center}
			\includegraphics[width=1.0\columnwidth,clip]{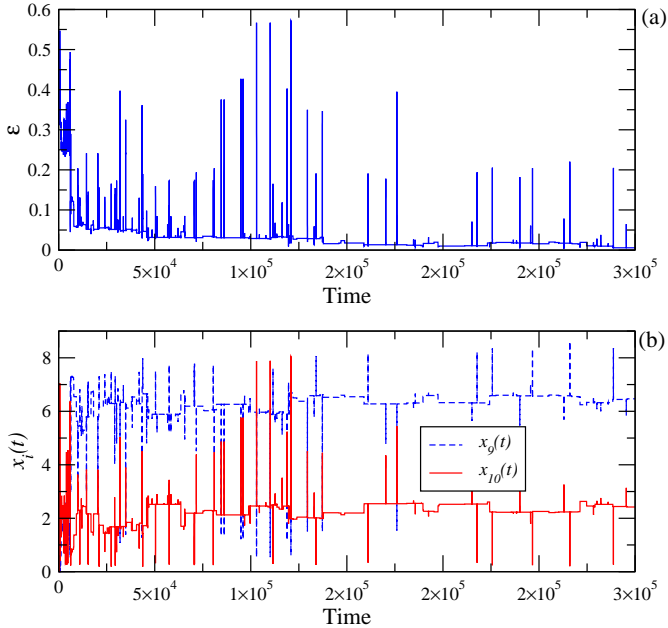} 
			\caption{(a) Error $\epsilon$ as a function of time. (b) Concentrations $x_{9}$ and $x_{10}$ of the two output nodes as a function of time. Target values $\hat{x}_{9} =6.5$ and $\hat{x}_{10}=2.5$. }
			\label{fig_evo_quimico} 
		\end{center}
	\end{figure}

	In fig. \ref{fig_evo_quimico} we present the time evolution for this system. Figure \ref{fig_evo_quimico}a shows the evolution of the error $\epsilon(t)$ as a function of time. We observe that this quantity decreases almost to zero at the end of the simulation. Note the strong variations during this process, signaling the strong effects of mutations on the functionality when the system has small error values. Figure \ref{fig_evo_quimico}b shows the evolutions of the concentrations for the target compounds (nodes $9$ and $10$). We can see that both concentrations can almost reach the target values. Finally, in fig. \ref{fig_nets} we present the initial and final networks for this example.

	We focus now on the dependence of the error $\epsilon$ on the noise intensity parameter $S$ and the delay $\eta$. We consider ensembles of $100$ networks with different random targets and initial conditions. In this study, networks have $N=10$ elements with $N_{out}=2$ output nodes. The target values  $\hat{x}_9$ and $\hat{x}_{10}$ are randomly chosen with uniform distributions between zero and one, and later normalized in order to satisfy that $\hat{x}_9 + \hat{x}_{10}= 0.9 I_1^{ext}/\gamma$. We consider again $I_1^{ext}=1$. The simulation have $10^5$ epochs.  
	
        \begin{figure}[!ht] 
		\begin{center}
			\includegraphics[width=1.0\columnwidth,clip]{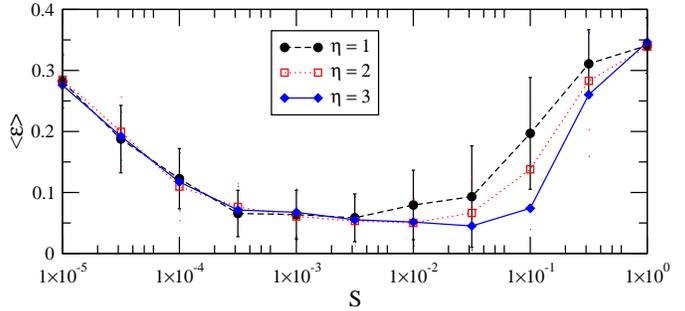} 
			\caption{(a) Mean error $\langle \epsilon \rangle$ as a function of the noise intensity parameter $S$ for three different values of delay $\eta$. Error bars indicated the dispersion of the distributions of the realizations for each ensemble.}
			\label{fig_quimica_estadisticas} 
		\end{center}
	\end{figure}

	Figure \ref{fig_quimica_estadisticas} shows the mean error $\langle \epsilon \rangle$ as a function of the noise intensity parameter $S$ for different values of $\eta$. We observe that the curves have similar behaviour for all three cases. The minimum mean values are also similar for the three curves, which indicates that the three proposed delay values have similar performance. We note, however, that for $\eta=3$ there is a broader range of noise intensities with small errors. We find that there is relatively large dispersion for the final error values inside each ensemble of networks. This shows that the difficulty of the random target patterns can be quite different among them. In conclusion, we observe that there is an optimum value of noise intensity at $S \approx 3.2 \times 10^{-2}$ that produces functional networks with smaller error. If the noise intensity is very low, the ratio of mutation is also low and there are not enough epochs in the simulation to find functional networks. On the contrary, large values of noise producs a large ratio and the system cannot find good solutions. 
	
	Note that both this system and the previous one, an error value of zero cannot be reached generally. In effect, these networks with ten nodes have $2^{N^2-N} \approx 10^{27}$ configurations, but the target concentrations have random real values. As a result, the learning capacity is limited to finding a good configuration that minimizes the error. Since the error $\epsilon$ is different from zero, the noise intensity does not vanish and random mutations are always present. In consequence, we can find large error values during the evolution even when the system is reaching configurations with small error.

	\section{Discussion and Conclusions}
	
	In this work, we have designed link evolution dynamics that can be expressed as a map where the link state is updated after fixed time intervals. The proposed dynamics are constructed following our conjecture of autonomous learning previously presented for systems that feature parameters with continuous time domain \cite{kaluza_autonomous}. We show through two different systems that the network architectures can evolve autonomously to produce a target pattern that minimizes an error function.
	
	The formulation we present in this manuscript has the same scheme of parameter updated through a map as our previous discrete time model shown in ref. \cite{kaluza_autonomous_discrete}. In effect, in both formulations we require that the time interval $\tau$ between updates be long enough to allow the system to reach a well-defined error value.
	
	Although our learning scheme can be seen just as one more optimization algorithm that is and even less efficient than others well known, our main goal is to ability to write a dynamical system for the whole process.  This is because the autonomous learning conjecture opens the possibility to create a system able to perform a prescribed task in a self-organized way without needing a tutor or a central processing unit. In effect, a dynamical system that contains not only the evolution laws of the system variables but also the learning mechanisms can be constructed as a physical system. For example, it is known that analog computers are electronic circuits especially designed in order to integrate systems of differential equations in real time. As a result, an electronic circuit that is able to learn by our proposed mechanisms is a concrete possibility. These characteristics are essential for applications where no external control or algorithmic evolutions can be implemented, such as biological and nano systems \cite{Mikhailov_libro_machines}.  
	
	 \acknowledgments
     
	Author acknowledges financial support from SeCTyP – UNCuyo (project M028 2017-2018) and from CONICET (PIP 11220150100013), Argentina.

\end{document}